\documentclass[12pt]{article}
\usepackage{changes}
\usepackage{multicol}
\usepackage{amssymb}
\usepackage{amsfonts,amssymb,amsmath,amsthm}
\usepackage{epsfig}
\usepackage{slashed}
\usepackage{graphicx}
\usepackage{subfig}
\usepackage{mathrsfs}
\usepackage{bbm}
\usepackage{cite}
\usepackage{enumerate}
\usepackage{slashbox}
\usepackage{color}
\usepackage[colorlinks,linkcolor=red,anchorcolor=red,citecolor=green]{hyperref}
\makeatletter 

\makeatletter

\newcommand{\Rmnum}[1]{\expandafter\@slowromancap\romannumeral #1@}
\makeatother

\begin{document}
\renewcommand{\thefootnote}{\fnsymbol{footnote}}
\begin{titlepage}

\vspace{10mm}
\begin{center}
{\Large\bf Noncommutative corrections to the minimal surface areas of the pure AdS spacetime and Schwarzschild-AdS black hole}
\vspace{10mm}

{{\large Zhang-Cheng Liu${}^{1,}$\footnote{\em E-mail address: zhangchenliuxt@gmail.com}
and Yan-Gang Miao${}^{1,2,}$\footnote{\em Corresponding author. E-mail address: miaoyg@nankai.edu.cn}}

\vspace{6mm}
${}^{1}${\normalsize \em School of Physics, Nankai University, Tianjin 300071, China}

\vspace{3mm}
${}^{2}${\normalsize \em Bethe Center for Theoretical Physics and Institute of Physics, University of Bonn, \\
Nussallee 12, 53115 Bonn, Germany}
}
\end{center}

\vspace{5mm}
\centerline{{\bf{Abstract}}}
\vspace{6mm}
Based on the perturbation expansion, we compute the noncommutative corrections to the minimal surface areas of the pure AdS spacetime and Schwarzschild-AdS black hole, where the noncommutative background is suitably constructed in terms of the Poincar\'e coordinate system. In particular, we find a reasonable tetrad with subtlety, which not only matches the metrics of the pure AdS spacetime and Schwarzschild-AdS black hole in the commutative case, but also makes the corrections real rather than complex in the noncommutative case.
For the pure AdS spacetime, the nocommutative effect is only a logarithmic term,  while for the Schwarzschild-AdS black hole, it contains a logarithmic contribution plus both a mass term and a noncommutative parameter related term. Furthermore, we show that the holographic entanglement entropy with noncommutativity obeys a relation which is similar to the first law of thermodynamics in the pure AdS spacetime.

\vspace{5mm}
\noindent
{\bf PACS Number(s)}: 02.30.Mv, 04.50.Kd, 04.60.-m

\vspace{5mm}
\noindent
{\bf Keywords}: Noncommutative geometry, holographic entanglement entropy

\end{titlepage}

\newpage
\renewcommand{\thefootnote}{\arabic{footnote}}
\setcounter{footnote}{0}
\setcounter{page}{2}
\pagenumbering{arabic}
\tableofcontents
\vspace{1cm}

\section{Introduction}
One way to understand quantum gravity follows the holographic correspondence. In terms of the open/closed string duality~\cite{JMM,Gubser:1998bc,Witten:1998qj}, the AdS/CFT correspondence
connects a $(d + 1)$-dimensional AdS space with a $d$-dimensional CFT which is located at
the boundary of the AdS space. Though it is hard to give a general proof for this kind
of dualities, some evidences give us specific ideas on how to connect the classical gravity with
a quantum gravity. One of the examples is the Ryu-Takayanagi (RT) formula~\cite{SRTT,Ryu:2006ef} in which the entanglement entropy of a CFT in subsystem A\footnote{The entanglement entropy of subsystem A is defined as the von Neumann entropy $S_A$ when the degrees of freedom in subsystem A's complement, subsystem B, are traced
	out. It describes how the two subsystems A and B are entangled or
	correlated each other.} is proportional to the minimal
surface area $\gamma_A$ in the AdS spacetime whose boundary is located at $\partial\gamma_A=\partial A$.\footnote{A closely related quantity is the black hole entropy which is proportional to the horizon area according to the Bekenstein-Hawking formula. One can refer to Refs. \cite{Frolov:1998vs,Susskind:1994sm,Callan:1994py} for the related
	discussions about the black hole entropy from the viewpoint of quantum field theory. The entanglement entropy computed from the RT formula is equivalent to the black hole entropy in
	certain cases, such as in the AdS black hole \cite{SRTT}  and in the black hole on a brane \cite{Emparan:2006ni}.} The proof of the formula sheds~\cite{ALJM,Dong:2016fnf, HCMH} light on understanding the AdS/CFT duality in the cosmic brane or 
the topological black hole background. In addition, the covariant entanglement entropy has
also been studied~\cite{VEHM}, where one can obtain more information about the time evolution of
the entanglement entropy. However, there are still many problems that should be solved before the
holographic dictionary is completed. Here we list some of the problems and the literature
in which some efforts have been made to try to deal with these problems.

\begin{itemize}
\item How to encode~\cite{DH} the bulk information from the field theory on the boundary when the information seems to be non-local?
\item How to extend~\cite{XD,LRMM,deBoer:2011wk,Miao:2014nxa,Fursaev:2006ih} the RT formula beyond the classical gravity? What are the general conditions that the formula should meet?
\item Whether can the AdS/CFT duality be extended~\cite{AStrom,KN,KN2,WLTT} to the de Sitter spacetime or flat spacetime besides the consideration of symmetries?
\item What happens~\cite{TFAL,SNS} in the dual theory if the quantum field is taken into account in the bulk?
\end{itemize}

Though many questions mentioned above have been studied on the side of gravity, the relevant questions on the side of quantum field theory
have not been dealt with. Moreover, the entanglement entropy
does not always meet the area law. In order to complete
the holographic dictionary, the noncommutativity of spacetimes should be considered. Here,
we briefly review some interesting models of noncommutative geometry and give their applications in quantum gravity. Among the attempts to quantize gravity, a natural thought
is to consider noncommutative spacetimes, early dubbed quantized spacetimes that could
be traced back to Snyder's pioneering work~\cite{HSS}. The revival of noncommutative spacetimes
about half a century after Snyder's work originated~\cite{NSEW} from the low energy effective field
theory of string theory. In Ref.~\cite{NSEW}, the Seiberg-Witten map establishes a connection between one
gauge theory on a noncommutative spacetime and another theory on the commutative spacetime. Based on
this map, one can investigate~\cite{Cacciatori:2002ib,Banados:2001xw} a noncommutative gravity by rewriting it as a gauge theory.
According to the coordinate coherent state formalism \cite{Smailagic:2003yb}, which records the noncommutativity
of coordinates by the spread of coherent states, the black holes in noncummutative spacetimes were constructed~\cite{Nicolini:2008aj,Nozari:2008rc,Rizzo:2006zb,Rahaman:2013gw,Nicolini:2005vd} if the point-like source was replaced with the smearing of objects.
One may also refer to the various attempts in the constructions of the noncommutiative gravity \cite{Harikumar:2006xf,Moffat:2000gr}, the noncommutative black holes  \cite{AJAS}, and the noncommutative quantum cosmology~\cite{Bastos:2007bg}.

Besides the AdS/CFT correspondence with noncommutativity, we have to study the holographic entanglement entropy with noncommutativity in order to understand the effects of non-locality on quantum entanglement.
While the entanglement entropy of a local field theory obeys the area law \cite{Srednicki:1993im,Holzhey:1994we}, which
shows the strong correlations among the states near the boundary, the area law is not necessarily true for a non-local field theory. For instance, the entanglement entropy of the non-local
scalar field theory obeys \cite{NT,WLTT} the volume law. Furthermore, it is of significance to study
the non-leading contributions to holographic entanglement entropy. On the one hand, the
non-leading contributions can extend the dictionary beyond the area term. For example, the
background entanglement from the one-loop quantum correction contributes~\cite{TFAL,SNS} extra
terms to entanglement entropy, and the noncommutative effect contributes~\cite{WFAK}
an extra divergent term to the holographic entanglement entropy in a noncommutative gauge theory. On the other hand, the information contained in the non-leading contributions is meaningful. The  non-leading contributions to entanglement entropy
encode~\cite{MPH} the universal terms which discriminate different phases.
And the higher derivative effect of gravity yields \cite{Miao:2014nxa}
the correct universal term for the generalized holographic entanglement entropy in the 4d CFT.

In the present work, we add a non-leading term to the surface area, where this term is induced by the noncommutative structure of spacetimes.
We find a specific tetrad and propose a noncommutative relation based on the Poincar\'e coordinate system.
Then, the corresponding Moyal product can be given, which is different from those already constructed~\cite{DMMR,MCAT,MRMR}.
Our treatment is a physically straightforward modification to the classical theory of gravity.
As it stems from the spacetime structure itself, the noncommutative geometry will show its significance to the holographic entanglement entropy on the side of quantum field theory.

The procedure goes as follows. We establish a noncommutative construction by using the Moyal product in the Poincar\'e coordinate system. Then, according to this construction, we compute the minimal surface areas of the pure AdS spacetime and Schwarzschild-AdS black hole. As the noncommutative parameter is much smaller than one, it is reasonable~\cite{YSLZ,NKJHL} to expand perturbatively a geodesic curve with respect to this parameter.
That is, it is explicit to study the noncommutative contributions to the minimal surface areas order by order. We find that the noncommutative geometry contributes a logarithmic divergent term for the pure AdS spacetime, where this term is proportional to the noncommutative parameter. And we make a similar discussion for the Schwarzschild-AdS black hole by the additional consideration of the black hole mass, and the result contains both a mass term and a noncommutative parameter\added{-}related term besides the logarithmic contribution. Furthermore, for the pure AdS spacetime with noncommutativity, we show that the noncommutative holographic entanglement entropy obeys a relation that is similar to the first law of thermodynamics.

The rest of the paper is organized as follows. In section 2, we compute the minimal surface area of the pure AdS spacetime with noncommutativity.  Then, we turn to the Schwarzschild-AdS black hole in section 3, where the perturbation of the mass parameter should be performed before the consideration of the noncommutative correction. In section 4, we consider the noncommutative stress tensor for the pure AdS spacetime  and analyze the thermodynamic property of the holographic entanglement entropy with noncommutativity.  Finally, we summarize our results in section 5.

\section{The minimal surface area of the pure AdS spacetime with noncommutativity}\label{sec2}

The metric of the (3+1)-dimensional AdS spacetime in the Poincar\'e coordinate system takes the following form,
\begin{equation}
ds^2=L^2\,\frac{-dt^2+dx^2+dy^2+dz^2}{z^2},
\end{equation}
where $L$ represents the AdS radius. Due to the rotational symmetry in the two models studied in this and the next sections, it is convenient to adopt the polar coordinates, ($\rho$, $\phi$), in the $(x, y)$ plane, and to rewrite the above metric to be
\begin{equation}
ds^2=L^2\,\frac{-dt^2+d\rho^2+\rho^2d\phi^2+dz^2}{z^2}.
\end{equation}

If one wants to rewrite the above metric by using tetrads, there exist many equivalent formulations of tetrads, i.e. any one of them gives the same metric. However, we find that the formulations of tetrads are of subtlety when the metric is generalized to a noncommutative spacetime. That is, we can write many equivalent formulations of tetrads in the commutative case, but most of them lead to a complex correction to minimal surface areas in the noncommutative case, which is usually regarded as an unphysical result. Fortunately, we find out such a tetrad that meets the requirement in the commutative spacetime and simultaneously gives a real noncommutative correction to minimal surface areas for the pure AdS spacetime and Schwarzschild-AdS black hole. This tetrad can be set to be
\begin{equation}
(k^a)_{\mu}=(l^a,n^a,m^a,w^a),
\end{equation}
where the four components are determined\footnote{For instance, we can give another tetrad which will give rise to a complex noncommutative correction to minimal surface areas,
\begin{equation*}
l^a=L\left(\frac{\delta^a_0}{z}\right),\qquad
n^a=L\frac{\delta^a_1}{z},  \qquad m^a=L\left(\frac{\rho\delta^a_2}{2z^2}+\rho\delta^a_3\right), \qquad w^a=L\left(\frac{\delta^a_2}{2z^2}+\delta^a_3\right).
\end{equation*}
It is equivalent to eq.~(\ref{tecom}) in the commutative spacetime.} in terms of the Kronecker delta as follows,
\begin{equation}
l^a=L\left(\frac{i\delta^a_0}{2z^2}+i\delta^a_2 \right),\qquad
n^a=L\frac{\delta^a_1}{z},  \qquad m^a=L\left(-i\rho\frac{\delta^a_0}{2z^2}+i\rho\delta^a_2\right), \qquad w^a=L\frac{\delta^a_3}{z}.\label{tecom}
\end{equation}
Therefore, we construct the metric by using the above tetrad as follows,
\begin{equation}
g_{\mu\nu}=\eta_{ab}k^a_\mu k^b_\nu,\label{tetradmetric}
\end{equation}
where $\eta_{ab}$ is defined by\footnote{For the unphysical tetrad mentioned above in footnote 3, the corresponding $\eta_{ab}$ takes the form,
\begin{equation*}
\eta_{ab}={
\left(\begin{array}{cccc}
 -1&0&0&0\\0&1&0&0\\ 0&0&0&1\\ 0&0&-1&0
\end{array}
\right)}.
\end{equation*}}
\begin{equation}
\eta_{ab}={
\left(\begin{array}{cccc}
 0&0&1+i&0\\0&1&0&0\\ 1-i&0&0&0\\ 0&0&0&1
\end{array}
\right)}.
\end{equation}

We introduce a new variable $q$ which is defined by $q\equiv 1/z$, so that $q\to \infty$ corresponds to the infinity of the AdS spacetime. In particular, such a transformation avoids the ambiguity when the inverse of a coordinate is set to be an operator, i.e. when an ordinary spacetime is generalized to a noncommutative one.
Now we are ready to propose the following noncommutative relation by using the commutator between the operator of $q$ and that of $\rho$,
\begin{equation}
\left[\hat{q},\hat{\rho}\right]=ih,\label{ncrelation}
\end{equation}
where $h$ is the noncommutative parameter that is dimensionless. 

Under the above noncommutative relation, the multiplication of functions in noncommutative geometry can be realized in terms of the Moyal product, see, for instance, the Moyal product of two functions with respect to two variables $u$ and $v$,
\begin{equation}
f(u,v)*g(u,v)=f(u,v)e^{i\frac{h}{2}(\stackrel{\leftarrow}{\partial_u}\stackrel{\rightarrow}{\partial_v}-\stackrel{\leftarrow}{\partial_v}\stackrel{\rightarrow}{\partial_u})}g(u,v),
\end{equation}
where $u=q$ and $v=\rho$ are specified in the present paper.
As a result, we can write the noncommutative metric from the ordinary (commutative) one, eq.~(\ref{tetradmetric}), as follows,
\begin{equation}
\bar{g}_{\mu\nu}=\eta_{ab}k^a_\mu*k^b_\nu,\label{moyalmetric}
\end{equation}
from which we derive the square of the line element with the noncommutativity up to the first order of $h$,
\begin{equation}
d\bar{s}^2= L^2\left[-q^2dt^2+hqdt d\phi+q^2d\rho^2+(q^2\rho^2-hq\rho) d\phi^2+\frac{dq^2}{q^2}\right].\label{ncmetric}
\end{equation}

According to the Poincar\'e coordinate system, the projection of the minimal surface into the plane depicted by the coordinates $\rho$ and $q^{-1}$ is a circle when $\phi$ is integrated from zero to $2\pi$, so it is convenient to introduce the following polar coordinate system, $(r,\theta)$, for computing the minimal surface area,
\begin{equation}
 \rho=r\cos\theta, \qquad q^{-1}=r\sin\theta,\label{cotrans}
 \end{equation}
where $\theta\in[\epsilon,\pi/2]$, $\epsilon$ is a regularization factor that is close to zero and associated with a lattice length of fields on the boundary, and $r$ represents the spacetime scale. 

Now we can easily write the minimal surface area with the noncommutative deformation by using eqs.~(\ref{ncmetric}) and (\ref{cotrans}) as follows,
\begin{equation}
A=2\pi L^2\int_\epsilon^{\frac{\pi}{2}}\frac{\sqrt{({\dot r}^2+r^2)(\cos^2\theta-h\sin\theta \cos\theta)}}{r\sin^2\theta}d\theta,\label{action1}
\end{equation}
where a dot stands for the differentiation with respect to $\theta$.
As the noncommutative parameter $h$ is much smaller than one, we expand the geodesic curve $r(\theta)$ with respect to it,
\begin{equation}
r(\theta)=r_0+h\bar{r}(\theta)+O(h^2),\label{1storder}
\end{equation}
where $r_0$ is the initial constant corresponding to the spatial scale and $\bar{r}(\theta)$ represents the first order noncommutative modification to the curve.
Because the term related to ${\dot r}^2$ is proportional to $h^2$,  see eq.~(\ref{1storder}), the minimal surface area, eq.~(\ref{action1}), reduces approximately to the following form if only the first order of $h$ is considered in the square root,
\begin{eqnarray}
A&=& 2\pi L^2\int_\epsilon^{\frac{\pi}{2}}\frac{\sqrt{\cos^2\theta-h\sin\theta \cos\theta}}{\sin^2\theta}d\theta \nonumber \\
&\equiv & A_0+\bar{A},\label{minisur}
\end{eqnarray}
where $A_0$ is the ordinary (commutative) minimal surface area and $\bar{A}$ is the first order noncommutative correction to $A_0$.
Making the Taylor expansion of eq.~(\ref{minisur}) with respect to $h$, we can easily obtain $A_0$ and $\bar{A}$, respectively,\footnote{According to the RT formula mentioned in Introduction, the undeformed holographic entanglement
	entropy ($A_0/(4G_N)$) corresponds to the entanglement entropy of the boundary field between the
	disk A and its complement region B because the spacetime keeps the rotating symmetry.}
\begin{equation}
A_0=2\pi L^2\left(\frac{1}{\sin\epsilon}-1\right),\label{a0}
\end{equation}
and
\begin{equation}
\bar{A}=\pi L^2 h\log\frac{\epsilon}{2},\label{1stcorr}
\end{equation}
where the latter can also be understood as the nonlocal contribution from the noncommutative spacetime.


In this section, the noncommutative effect is manifested through the Moyal product of
our tetrad (eq.~(\ref{tecom})), see eq.~(\ref{moyalmetric}). We calculate the noncommutative minimal surface
by using the deformed metric eq.~(\ref{ncmetric}).
From eq.~(\ref{1stcorr}), we see that the noncommutative effect gives a logarithmic divergent term with a suppression factor $h$. Because $\bar{A}$ is minus, the noncommutative effect decreases the minimal surface area of the pure AdS spacetime. 

\section{The minimal surface area of the Schwarzschild-AdS black hole with noncommutativity}\label{sec3}
We deal with the Schwarzschild-AdS black hole in two steps. In the first step, we compute its minimal surface area by regarding~\cite{HLSJ,YSLZ} the mass of the Schwarzschild black hole, $M$, as a perturbative parameter in the pure AdS spacetime. Then, following the method we applied in the above section, we derive in the second step the noncommutative correction to the minimal surface area of the Schwarzschild-AdS black hole. That is, based on the pure AdS spacetime, our result contains the contributions from the pure AdS spacetime together with the mass correction and the noncommutative parameter correction.

The metric of the (3+1)-dimensional Schwarzschild-AdS black hole takes the form~\cite{HLSJ},
\begin{equation}
ds^2=\frac{L^2}{z^2}\left(-f(z)dt^2+\frac{dz^2}{f(z)}+d\rho^2+\rho^2d\phi^2\right),
\end{equation}
where $f(z)$ that is associated with the black hole mass $M$ reads
\begin{equation}
f(z)=1-M z^3.
\end{equation}
As just mentioned above, $M$ is treated as a perturbative parameter under the condition $M z^3\ll 1$, where such a reparametrization has been made that $M z^3$ is dimensionless.

In the first step, 
we give the minimal surface area of the Schwarzschild-AdS black hole in the polar coordinate $(r,\theta)$,
\begin{align}
{\cal A}_0= 2\pi L^2\int_\epsilon^{\frac{\pi}{2}}&\left[{r^2+\dot r^2+Mr^3\sin^3\theta\left(1+Mr^3\sin^3\theta\right)\left(r^2\cos^2\theta+r\dot r\sin2\theta+\dot  r^2\sin^2\theta\right)}\right]^{1/2}\nonumber \\
& \times \frac{\cos\theta}{r\sin^2\theta}d\theta.\label{mArea}
\end{align}
By expanding $r(\theta)$ with respect to $M$, we have
\begin{equation}
r(\theta)=l+Ma(\theta)+O(M^2),\label{mcurve}
\end{equation}
where $l$ is a constant that is associated with a spacial scale and satisfies the inequality $M l^3\ll 1$. Substituting eq.~({\ref{mcurve}}) into eq.~({\ref{mArea}}), we then derive
the leading contribution up to the first\added{-}order of mass $M$, 
\begin{eqnarray}
{\cal A}_0&=&2\pi L^2\int_\epsilon^{\frac{\pi}{2}}\left(1+\frac{1}{2}{Ml^3\sin^3\theta\cos^2\theta}\right)\frac{\cos\theta}{\sin^2\theta}d\theta\nonumber \\
&\equiv &A_0+A_{\rm M},
\end{eqnarray}
where $A_0$ is the minimal surface area of the pure AdS spacetime, see eq.~(\ref{a0}), and $A_{\rm M}$ represents the mass correction to $A_0$,
\begin{equation}
A_{\rm M}=\frac{1}{4}{\pi L^2 Ml^3}.
\end{equation}
As a result, we work out the minimal surface area of the Schwarzschild-AdS black hole, ${\cal A}_0$, in the approximation of the first\added{-}order of $M$.

Now we calculate the noncommutative correction to ${\cal A}_0$ in the second step by following the way utilized in the above section. 
Combining eq.~(\ref{mArea}) together with eq.~(\ref{ncmetric}), we write the minimal surface area of the Schwarzschild-AdS black hole with noncommutativity,
\begin{align}
{\cal A}=2\pi L^2\int_\epsilon^{\frac{\pi}{2}}& \left[r^2+\dot r^2+Mr^3\sin^3\theta(1+Mr^3\sin^3\theta)(r^2\cos^2\theta+r\dot r\sin2\theta+\dot r^2\sin^2\theta)\right]^{1/2}\nonumber \\
& \times \left(\cos^2\theta-h\sin\theta\cos\theta\right)^{1/2}\frac{d\theta}{r\sin^2\theta}.\label{mncarea}
\end{align}
Defining
\begin{equation}
{\cal A}\equiv{\cal A}_0+\bar{\cal A},
\end{equation}
and substituting both eq.~(\ref{1storder}) and eq.~(\ref{mcurve}) into eq.~(\ref{mncarea}), we derive
the contribution from the noncommutative modification up to the first order of $h$,
\begin{eqnarray}
\bar{{\cal A}}&=&-\pi L^2h\int_\epsilon^{\frac{\pi}{2}}\left(1+\frac{1}{2}Ml^3\sin^3\theta\cos^2\theta\right)\frac{d\theta}{\sin\theta}\nonumber \\
&=&\pi L^2 h\log\frac{\epsilon}{2}-\frac{\pi}{32}\,{\pi L^2Ml^3h}.\label{nccorrentr}
\end{eqnarray}
When the mass parameter is set to be zero, the above result goes back to that of the pure AdS situation, see the first term of eq.~(\ref{nccorrentr}) or eq.~(\ref{1stcorr}). On the other hand, besides the noncommutative correction from the noncommutative AdS spacetime, the noncommutativity also modifies the contribution from the mass term of the black hole, see the second term of eq.~(\ref{nccorrentr}).
\section{Thermodynamic property of holographic entanglement entropy with noncommutativity}
To further analyze the noncommutative effects on the thermodynamic property of holographic entanglement entropy, we study the stress tensor under the noncommutative deformation. For simplicity, the following analysis is based on the deformed spacetime eq.~(\ref{ncmetric}), where the noncommutative effects associated with higher orders of $h$ are ignored. In this way, we regard the noncommutative Minkowski spacetime as the pure AdS spacetime associated with matters. We note that the holographic entanglement entropy is corrected noncommutatively as given in section 2 and section 3 when the background metric is deformed by eq.~(\ref{ncmetric}), which means that the holographic entanglement entropy remains unchanged whether the noncommutative effect is regarded as matter or not.
The reason lies in the fact that the holographic entanglement entropy is only sensitive to the background metric.

We start with the gravitational action of the noncommutative $(3+1)$-dimensional pure Ads spacetime \cite{BVKP,Hartman:2018tkw} with matter,
\begin{equation}
S=\frac{1}{16\pi G_N}\int_{\mathcal{M}}d^4x\sqrt{-g}\left({R}+\frac{6}{L^2}\right)+S_{\rm matter}-\frac{1}{8\pi G_N}\int_{\partial\mathcal{M}}d^3x\sqrt{-\gamma}\,\Theta+\frac{1}{8\pi G_N}S_{\rm ct}(\gamma_{\mu\nu}),\label{action}
\end{equation}
where ${R}$ is the Ricci scalar of the spacetime, $S_{\rm matter}$ is the matter action equivalent to the noncommutative contribution, $\gamma$ is the boundary metric,  $\Theta$ is the trace of  the extrinsic curvature $\Theta_{\mu\nu}=\frac{1}{2}(\nabla_\mu n_\nu+\nabla_\nu n_\mu)$ with $n_\mu$ the outward pointing normal vector to the boundary $\partial\mathcal{M}$, and $S_{\rm ct}$ is the counterterm action. To get a finite stress tensor on the boundary $\partial\mathcal{M}$, the counterterm action $S_{\rm ct}$ is chosen to be a covariant function,
\begin{equation}
S_{\rm ct}=-\frac{2}{L}\int_{\partial M}\sqrt{-\gamma}\left( 1-\frac{L^2}{4}{{R}}\right)d^3x.\label{counter}
\end{equation}
Next, we compute the stress tensor,
\begin{equation}
T_{\mu\nu}^{\rm grav}=-\frac{1}{8\pi G_N}\left(\Theta_{\mu\nu}- \Theta\gamma_{\mu\nu}+\frac{2}{L}\gamma_{\mu\nu}-L {G}_{\mu\nu}\right),
\end{equation}
where ${G}_{\mu\nu}={{R}}_{\mu\nu}-\frac{1}{2}{{R}}\gamma_{\mu\nu}$ is the Einstein tensor associated with $\gamma_{\mu\nu}$.
Substituting  eq.~(\ref{ncmetric}) into the above equation, we obtain the energy component,
\begin{equation}\label{stt1}
-8\pi G_NT_{00}^{\rm grav}=\frac{h L q}{2\rho}+\frac{L h^2}{4\rho^2}+O(h^3).
\end{equation}
We find from this component that the leading contribution without noncommutative corrections is vanishing,  which coincides with~\cite{BVKP} the usual result in the pure $AdS_4$ spacetime. The first term in the right-hand side of eq.~(\ref{stt1}) is divergent when we take $q\to\infty$ at the boundary, but there are no covariant ways to cancel it because the noncommutative correction is coordinate-dependent. In order to cancel the divergent term, we introduce a coordinate-dependent counterterm,
\begin{equation}
S_{\rm hct}=\frac{h}{2L}\int_{\partial M}\sqrt{-\gamma}\left(\frac{1}{\rho q}\right)d^3x,
\end{equation}
and add it to the action eq.~(\ref{action}). Thus, we obtain the finite result,
\begin{equation}
-8\pi G_N{\cal T}_{00}^{\rm grav}=\frac{L h^2}{4\rho^2}.\label{henergy}
\end{equation}

From eq.~(\ref{henergy}),
we can see that the noncommutative correction to the stress tensor is of order $h^2$ and depends only on the radial coordinate. As the stress tensor is divergent around $\rho=0$, we have to regularize it by replacing the vanishing radial coordinate with a cutoff $\epsilon'$, see also footnote 6 below.

In ref.~\cite{Bhattacharya:2012mi} an analogous relation with the first law of thermodynamics is given for a subsystem $A$ when the system is excited,
\begin{equation}
    T_{\rm ent}\cdot \Delta S_A=\Delta E_A,\label{floee}
\end{equation}
where $\Delta S_A$ measures how much $S_A$ (entanglement entropy) is increased in the exited state compared with the ground
state of the CFT on the boundary $\partial\mathcal{M}$, and $\Delta E_A$ is the increased amount of energy in the subsystem $A$. Note that
$T_{\rm ent}$ means the effective temperature  or the so-called entanglement temperature, which is proportional to the inverse of the length scale of subsystem $A$.

For the pure AdS spacetime with noncommutativity as discussed in section 2, we obtain the corresponding increased energy of subsystem $A$ by considering eq.~(\ref{henergy}),
\begin{equation}
    \Delta E'_A=\int d^2x {\cal T}_{00}^{\rm grav}=\int_{\epsilon'}^{l}2\pi\rho{\cal T}_{00}^{\rm grav}d\rho =\frac{Lh^2}{16G_N}\log\frac{\epsilon'}{l},\label{denergy}
\end{equation}
where we have regularized the stress tensor around the coordinate origin.\footnote{Note $\epsilon'\equiv\epsilon l$, where $l$ is the length scale of subsystem $A$, as to $\epsilon$, it is dimensionless, see eq.~(\ref{1stcorr}).} On the other hand, here $\Delta S'_A$ is just $\bar{A}$ (see eq.~(\ref{1stcorr})) divided by $4G_N$. Considering eq.~(\ref{denergy}) and eq.~(\ref{1stcorr}), we obtain the relation,
\begin{equation}
T'_{\rm ent}\cdot \Delta S'_A=\Delta E'_A+\frac{Lh^2}{16G_N}\log\frac{1}{2},\label{nfloee}
\end{equation}
where $T'_{\rm ent}\equiv\frac{h}{4\pi L}$, which plays a similar role to $T_{\rm ent}$ of eq.~(\ref{floee}). We notice that eq.~(\ref{nfloee}) is similar to eq.~(\ref{floee}) and the second term on its right-hand side is the subleading term that is originated from the noncommutativity.

According to eq.~(\ref{nfloee}), we find that the noncommutative correction of holographic entanglement entropy also obeys an analogous relation to the first law of thermodynamics. Comparing eq.~(\ref{nfloee}) with eq.~(\ref{floee}),
we believe that the noncommutative correction of holographic entanglement entropy is related to the increased amount of entanglement entropy from a certain excited state~\cite{Blanco:2013joa}.

\section{Summary}\label{sec4}
In this paper\added{,} we point out that the noncommutative generalization of the minimal surface areas is nontrivial. At the commutative level, there exist many equivalent formulations of tetrads that give rise to the same metric. However, at the noncommutative level, most of them lead to complex noncommutative corrections to the minimal surface areas, which is unphysical. Therefore, the construction of a specific tetrad is of subtlety. Fortunately, we have found such a tetrad and obtained the real noncommutative corrections to the minimal surface areas for the pure AdS spacetime and Schwarzschild-AdS black hole.

The RT formula shows the relation between the entanglement entropy $S_{A'}$ of conformal fields and the minimal surface area in Einstein gravity\cite{SRTT},
\begin{equation}
S_{A'}=\frac{\rm Area}{4G_N}.
\end{equation}
While the leading term of the holographic entanglement entropy is not altered by the noncommutativity of spacetimes, we observe that an extra divergent term is induced by the noncommutative geometry.
The noncommutative spacetime we consider in the present paper may be interpreted as the perturbation of matters. This leads to the result that the disturbance of the stress tensor at the boundary corresponds to the perturbed state from the vacuum state. Eq.~(\ref{nfloee}) shows that the holographic entanglement entropy with noncommutativity satisfies a similar relation to the first law of thermodynamics. While we do not have a direct interpretation of the holographic entanglement entropy with noncommutativity in the boundary theory, we believe that it corresponds to the entanglement entropy of excited states. Here we only consider the noncommutative holographic duality in the pure AdS spacetime. It is a nontrivial attempt to study the noncommutative holographic correspondence when spacetime is added with extra matters.

\section*{Acknowledgments}
Z-CL would like to thank X. Hao, Y. Sun, and Z.-M. Xu for helpful discussions.
Y-GM would like to thank H.P. Nilles for the warm hospitality during his stay at University of Bonn, and acknowledges the financial support from the Alexander von Humboldt Foundation under a renewed research program and from the National Natural Science Foundation of China under grant Nos. 11675081 and 12175108. In addition, the authors would like to thank the anonymous referees for the helpful comments that improve this work greatly.


\begin{thebibliography}{99}

\bibitem{JMM} J.M. Maldacena, {\em The large N limit of superconformal field theories and supergravity}, Int. J. Theor. Phys. {\bf 38}, 1113 (1999) [arXiv:hep-th/9711200].

\bibitem{Gubser:1998bc}
S.S.~Gubser, I.R.~Klebanov, and A.M.~Polyakov,
{\em Gauge theory correlators from noncritical string theory},
Phys. Lett. B {\bf 428}, 105 (1998)
[arXiv:hep-th/9802109 [hep-th]].

\bibitem{Witten:1998qj}
E.~Witten,
{\em Anti-de Sitter space and holography},
Adv. Theor. Math. Phys. \textbf{2}, 253 (1998)
[arXiv:hep-th/9802150 [hep-th]].

\bibitem{SRTT} S. Ryu and T. Takayanagi, {\em Holographic derivation of entanglement entropy from AdS/CFT}, Phys. Rev. Lett. {\bf 96}, 181602 (2006) [arXiv:hep-th/0603001].

\bibitem{Ryu:2006ef}
S.~Ryu and T.~Takayanagi,
{\em Aspects of holographic entanglement entropy},
JHEP \textbf{08}, 045 (2006)
[arXiv:hep-th/0605073 [hep-th]].

\bibitem{Frolov:1998vs}
V.P.~Frolov and D.V.~Fursaev,
{\em Thermal fields, entropy, and black holes},
Class. Quant. Grav. \textbf{15}, 2041 (1998)
[arXiv:hep-th/9802010 [hep-th]].

\bibitem{Susskind:1994sm}
L.~Susskind and J.~Uglum,
{\em Black hole entropy in canonical quantum gravity and superstring theory},
Phys. Rev. D \textbf{50}, 2700 (1994)
[arXiv:hep-th/9401070 [hep-th]].

\bibitem{Callan:1994py}
C.~Callan and F.~Wilczek,
{\em On geometric entropy},
Phys. Lett. B \textbf{333}, 55 (1994)
[arXiv:hep-th/9401072 [hep-th]].

\bibitem{Emparan:2006ni}
R.~Emparan,
{\em Black hole entropy as entanglement entropy: A holographic derivation},
JHEP \textbf{06}, 012 (2006)
[arXiv:hep-th/0603081 [hep-th]].

\bibitem{ALJM} A. Lewkowycz and J. Maldacena, {\em Generalized gravitational entropy}, JHEP {\bf 08}, 090 (2013) [arXiv:1304.4926 [hep-th]].

\bibitem{Dong:2016fnf}
X.~Dong,
{\em The gravity dual of Renyi entropy},
Nature Commun. \textbf{7}, 12472 (2016)
[arXiv:1601.06788 [hep-th]].

\bibitem{HCMH} H. Casini, M. Huerta, and R.C. Myers, {\em Towards a derivation of holographic entanglement entropy}, JHEP {\bf 05}, 036 (2011) [arXiv:1102.0440 [hep-th]].

\bibitem{VEHM} V.E. Hubeny, M. Rangamani, and T. Takayanagi, {\em A covariant holographic entanglement entropy proposal}, JHEP {\bf 07}, 062 (2007) [arXiv:0705.0016 [hep-th]].

\bibitem{DH} D. Harlow, {\em TASI lectures on the emergence of the bulk physics in AdS/CFT}, arXiv:1802.01040 [hep-th].

\bibitem{XD} X. Dong, {\em Holographic entanglement entropy for general higher derivative gravity}, JHEP {\bf 01}, 044 (2014) [arXiv:1310.5713 [hep-th]].

\bibitem{Miao:2014nxa}
R.-X.~Miao and W.-z.~Guo,
{\em Holographic entanglement entropy for the most general higher derivative gravity},
JHEP \textbf{08}, 031 (2015)
[arXiv:1411.5579 [hep-th]].

\bibitem{LRMM} L.-Y. Hung, R.C. Myers, and M. Smolkin, {\em On holographic entanglement entropy and higher curvature gravity}, JHEP {\bf 04}, 025 (2011) [arXiv:1101.5813 [hep-th]].

\bibitem{Fursaev:2006ih}
D.V. Fursaev,
{\em Proof of the holographic formula for entanglement entropy},
JHEP \textbf{09}, 018 (2006)
[arXiv:hep-th/0606184 [hep-th]].

\bibitem{deBoer:2011wk}
J. de Boer, M. Kulaxizi, and A. Parnachev,
{\em Holographic entanglement entropy in Lovelock gravities},
JHEP \textbf{07}, 109 (2011)
[arXiv:1101.5781 [hep-th]].

\bibitem{AStrom}
A. Strominger, {\em The dS/CFT correspondence}, JHEP {\bf 10}, 034 (2001) [arXiv:hep-th/0106113].


\bibitem{KN} K. Narayan, {\em Extremal surfaces in de Sitter spacetime}, Phys. Rev. D {\bf 91}, 126011 (2015) [arXiv:1501.03019 [hep-th]].

\bibitem{KN2} K. Narayan, {\em De Sitter space and extremal surfaces for spheres}, Phys. Lett. B {\bf 753}, 308 (2016) [arXiv:1504.07430 [hep-th]].

\bibitem{WLTT} W. Li and T. Takayanagi, {\em Holography and entanglement in flat spacetime}, Phys. Rev. Lett. {\bf 106}, 141301 (2011) [arXiv:1010.3700 [hep-th]].

\bibitem{TFAL} T. Faulkner, A. Lewkowycz, and J. Maldacena, {\em Quantum corrections to holographic entanglement entropy}, JHEP {\bf 11}, 074 (2013) [arXiv:1307.2892 [hep-th]].

\bibitem{SNS} S.N. Solodukhin, {\em Entanglement entropy of black holes}, Living Rev. Relativity {\bf 14}, 8 (2011) [arXiv:1104.3712 [hep-th]].

\bibitem{HSS} H.S. Snyder, {\em Quantized space-time}, Phys. Rev. {\bf 71}, 38 (1947).

\bibitem{NSEW} N. Seiberg and E. Witten, {\em String theory and noncommutative geometry}, JHEP {\bf 09}, 032 (1999) [arXiv:hep-th/9908142].

\bibitem{Cacciatori:2002ib}
S. Cacciatori, A.H. Chamseddine, D. Klemm, L. Martucci, W.A. Sabra, and D. Zanon,
{\em Noncommutative gravity in two dimensions},
Class. Quant. Grav. \textbf{19}, 4029 (2002)
[arXiv:hep-th/0203038 [hep-th]].

\bibitem{Banados:2001xw}
M. Banados, O. Chandia, N.E. Grandi, F.A. Schaposnik, and G.A. Silva,
{\em Three-dimensional noncommutative gravity},
Phys. Rev. D \textbf{64}, 084012 (2001)
[arXiv:hep-th/0104264 [hep-th]].

\bibitem{Smailagic:2003yb}
A.~Smailagic and E.~Spallucci,
{\em Feynman path integral on the noncommutative plane},
J. Phys. A \textbf{36}, L467 (2003)
[arXiv:hep-th/0307217 [hep-th]].

\bibitem{Nicolini:2008aj}
P.~Nicolini,
{\em Noncommutative black holes, the final appeal to quantum gravity: A review},
Int. J. Mod. Phys. A \textbf{24}, 1229 (2009)
[arXiv:0807.1939 [hep-th]].

\bibitem{Nicolini:2005vd}
P.~Nicolini, A.~Smailagic, and E.~Spallucci,
{\em Noncommutative geometry inspired Schwarzschild black hole},
Phys. Lett. B \textbf{632}, 547 (2006)
[arXiv:gr-qc/0510112 [gr-qc]].

\bibitem{Rizzo:2006zb}
T.G.~Rizzo,
{\em Noncommutative inspired black holes in extra dimensions},
JHEP \textbf{09}, 021 (2006)
[arXiv:hep-ph/0606051 [hep-ph]].

\bibitem{Nozari:2008rc}
K. Nozari and S.H. Mehdipour,
{\em Hawking radiation as quantum tunneling from noncommutative Schwarzschild black hole},
Class. Quant. Grav. \textbf{25}, 175015 (2008)
[arXiv:0801.4074 [gr-qc]].

\bibitem{Rahaman:2013gw}
F.~Rahaman, P.K.F.~Kuhfittig, B.C.~Bhui, M.~Rahaman, S.~Ray, and U.F.~Mondal,
{\em BTZ black holes inspired by noncommutative geometry},
Phys. Rev. D \textbf{87}, 084014 (2013)
[arXiv:1301.4217 [gr-qc]].

\bibitem{Harikumar:2006xf}
E.~Harikumar and V.O.~Rivelles,
{\em Noncommutative gravity},
Class. Quant. Grav. \textbf{23}, 7551 (2006)
[arXiv:hep-th/0607115 [hep-th]].

\bibitem{Moffat:2000gr}
J.W.~Moffat,
{\em Noncommutative quantum gravity},
Phys. Lett. B \textbf{491}, 345 (2000)
[arXiv:hep-th/0007181 [hep-th]].



\bibitem{AJAS} T. Juri\'c and A. Samsarov, {\em Entanglement entropy renormalization for the NC scalar field coupled to classical BTZ geometry}, Phys. Rev. D {\bf 93}, 104033 (2016) [arXiv:1602.01488 [hep-th]].

\bibitem{Bastos:2007bg}
C.~Bastos, O.~Bertolami, N.C. Dias, and J.N. Prata,
{\em Phase-space noncommutative quantum cosmology},
Phys. Rev. D \textbf{78}, 023516 (2008)
[arXiv:0712.4122 [gr-qc]].

\bibitem{Srednicki:1993im}
M.~Srednicki,
{\em Entropy and area},
Phys. Rev. Lett. \textbf{71}, 666 (1993)
[arXiv:hep-th/9303048 [hep-th]].

\bibitem{Holzhey:1994we}
C.~Holzhey, F.~Larsen, and F.~Wilczek,
{\em Geometric and renormalized entropy in conformal field theory},
Nucl. Phys. B \textbf{424}, 443 (1994)
[arXiv:hep-th/9403108 [hep-th]].

\bibitem{NT} N. Shiba and T. Takayanagi, {\em Volume law for the entanglement entropy in non-local QFTs}, JHEP {\bf 02}, 033 (2014) [arXiv:1311.1643 [hep-th]].

\bibitem{WFAK} W. Fischler, A. Kundu, and S. Kundu, {\em Holographic entanglement in a noncommutative gauge theory}, JHEP {\bf 01}, 37 (2014) [arXiv:1307.2932 [hep-th]].

\bibitem{MPH} M.P. Hertzberg, {\em Entanglement entropy in scalar field theory}, J. Phys. A {\bf 46}, 015402 (2013) [arXiv:1209.4646 [hep-th]].

\bibitem{DMMR} D. Momeni, M. Raza, and R. Myrzakulov, {\em Holographic entanglement entropy for noncommutative anti-de Sitter space}, Mod. Phys. Lett. A {\bf 31}, 1650073 (2016) [arXiv:1504.00106 [hep-th]].

\bibitem{MCAT} M. Chaichian, A. Tureanu, R.B. Zhang, and X. Zhang, {\em Riemannian geometry of noncommutative surfaces}, J. Math. Phys. {\bf 49}, 073511 (2008) [arXiv:hep-th/0612128].

\bibitem{MRMR} M.R. Mbonye and R.L. Mallett, {\em Gravitational perturbations of a radiating
	spacetime},  Found. Phys. {\bf 30}, 747 (2000) [arXiv:gr-qc/0010006].

\bibitem{YSLZ} Y. Sun and L. Zhao, {\em Holographic entanglement entropies for Schwarzschild and Reisner-Nordstr\"om black holes in asymptotically Minkowski spacetimes}, Phys. Rev. D {\bf 95}, 086014 (2017) [arXiv:1611.06442 [hep-th]].

\bibitem{NKJHL} N. Kim and J.H. Lee,  {\em Time-evolution of the holographic entanglement entropy and metric perturbations}, J. Korean Phys. Soc. {\bf 69}, 623 (2016) [arXiv:1512.02816 [hep-th]].

\bibitem{HLSJ} H. Liu and S.J. Suh, {\em Entanglement growth during thermalization in holographic systems}, Phys. Rev. D {\bf 89}, 066012 (2014) [arXiv:1311.1200 [hep-th]].

\bibitem{BVKP} V. Balasubramanian and P. Kraus, {\em A stress tensor for anti-de Sitter gravity}, Commun. Math. Phys. {\bf 208}, 413 (1999).

\bibitem{Hartman:2018tkw}
T.~Hartman, J.~Kruthoff, E.~Shaghoulian, and A.~Tajdini,
{\em Holography at finite cutoff with a $T^2$ deformation},
JHEP \textbf{03}, 004 (2019)
[arXiv:1807.11401 [hep-th]].

\bibitem{Bhattacharya:2012mi}
J.~Bhattacharya, M.~Nozaki, T.~Takayanagi, and T.~Ugajin,
{\em Thermodynamical property of entanglement entropy for excited states},
Phys. Rev. Lett. \textbf{110}, 091602 (2013) 
[arXiv:1212.1164 [hep-th]].

\bibitem{Blanco:2013joa}
D.D.~Blanco, H.~Casini, L.-Y.~Hung, and R.C.~Myers,
{\em Relative entropy and holography},
JHEP \textbf{08}, 060 (2013)
[arXiv:1305.3182 [hep-th]].


\end{thebibliography}
\end {document}